\documentclass[a4paper,fleqn,usenatbib]{mnras}

\usepackage{newtxtext,newtxmath}
\usepackage[T1]{fontenc}
\usepackage{ae,aecompl}

\usepackage{graphicx}	% Including figure files
\usepackage{amsmath}	% Advanced maths commands
\usepackage{amssymb}	% Extra maths symbols

\usepackage{graphicx,lscape,amssymb,amsmath,url,placeins,mathrsfs,float}

\def\aap{A\hbox{\rm \&}A} 
\def\aaps{A\hbox{\rm \&}A Suppl.}  \def\aj{AJ} \def\ao{Appl.\ Opt.}
  \def\apj{ApJ} \def\apjl{ApJ}
 \def\apjs{ApJS} 
\def\araa{ARA\hbox{\rm \&}A}

\def\mnras{MNRAS} \def\nat{Nat}

\title[IRX-$\beta$ relation at high-$z$]{A direct calibration of the
  IRX-$\bmath{\beta}$ relation in Lyman-break Galaxies at
  $\bmath{z}$=3--5}

\author[M. Koprowski et al.]
{M.~P.~Koprowski$^{1}$\thanks{E-mail: m.koprowski@herts.ac.uk},
K.~E.~K.~Coppin$^{1}$, 
J.~E.~Geach$^{1}$,
R.~J.~McLure$^{2}$,
O.~Almaini$^3$,
\newauthor
A.~W.~Blain$^4$,
M.~Bremer$^5$,
N.~Bourne$^2$,
S.~C.~Chapman$^6$,
C.~J.~Conselice$^7$,
J.~S.~Dunlop$^2$,
\newauthor
D.~Farrah$^8$,
W.~Hartley$^9$
A.~Karim$^{10}$,
K.~K.~Knudsen$^{11}$,
M.~J.~Micha{\l}owski$^{12}$,
\newauthor
D.~Scott$^{13}$,
C.~Simpson$^{14}$,
D.~J.~B.~Smith$^{1}$,
P.~P.~van der Werf$^{15}$,
\\
$^1$Centre for Astrophysics Research, School of Physics, Astronomy and Mathematics, University of Hertfordshire, College Lane, Hatfield AL10 9AB, UK\\
$^2$Institute for Astronomy, University of Edinburgh, Royal Observatory, Edinburgh EH9 3HJ, UK\\
$^3$School of Physics and Astronomy, University of Nottingham, University Park, Nottingham NG7 2RD, UK\\
$^4$Physics \& Astronomy, University of Leicester, 1 University Road, Leicester LE1 7RH, UK\\ 
$^5$Astrophysics Group, School of Physics, University of Bristol, Tyndall Avenue, Bristol BS8 1TL, UK\\
$^6$Department of Physics and Atmospheric Science, Dalhousie University, Halifax, NS B3H 4R2, Canada\\
$^7$University of Nottingham, School of Physics \& Astronomy, Nottingham, NG7 2RD, UK\\
$^8$Department of Physics, Virginia Tech, Blacksburg, VA 24061, USA\\
$^9$Department of Physics and Astronomy, University College London, London, WC1E 6BT, UK\\
$^{10}$Argelander-Institut f\"ur Astronomie, Universit\"at Bonn, Auf dem H\"ugel 71, D-53121 Bonn, Germany\\
$^{11}$Department of Space, Earth and Environment, Chalmers University of Technology, Onsala Space Observatory, SE-43992 Onsala, Sweden\\
$^{12}$Astronomical Observatory Institute, Faculty of Physics, Adam Mickiewicz University, ul.~S{\l}oneczna 36, 60-286 Pozna{\'n}, Poland\\
$^{13}$Department of Physics and Astronomy, 6224 Agricultural Road, University of British Columbia, Vancouver V6T 1Z1, Canada\\
$^{14}$Gemini Observatory, Northern Operations Center, 670 N. A'ohuku Place, Hilo, HI 96720, USA\\
$^{15}$Leiden Observatory, Leiden University, P.O. Box 9513, NL-2300 RA Leiden, The Netherlands\\
}

\date{Accepted XXX. Received YYY; in original form ZZZ}

\pubyear{2017}

\begin{document}
\label{firstpage}
\pagerange{\pageref{firstpage}--\pageref{lastpage}}
\maketitle

\begin{abstract}
We use a sample of 4178 Lyman break galaxies (LBGs) at $z\simeq3$, 4
and 5 in the UKIRT Infrared Deep Sky Survey (UKIDSS) Ultra Deep Survey
(UDS) field to investigate the relationship between the observed slope
of the stellar continuum emission in the ultraviolet, $\beta$, and the
thermal dust emission, as quantified via the so-called `infrared
excess' (${\rm IRX}\equiv L_{\rm IR}/L_{\rm UV}$). Through a stacking
analysis we directly measure the 850$\mu$m flux density of LBGs in our
deep (0.9\,mJy) James Clerk Maxwell Telescope (JCMT) SCUBA-2 850-${\rm
  \mu m}$ map, as well as deep public {\it Herschel}/SPIRE 250-, 350-
and 500-${\rm \mu m}$ imaging. We establish functional forms for the
IRX-$\beta$ relation to $z\sim 5$, confirming that there is no
significant redshift evolution of the relation and that the resulting
average IRX-$\beta$ curve is consistent with a Calzetti-like
attenuation law. We compare our results with recent work in the
literature, finding that discrepancies in the slope of the IRX-$\beta$
relation are driven by biases in the methodology used to determine the
ultraviolet slopes. Consistent results are found when IRX-$\beta$ is
evaluated by stacking in bins of stellar mass, and we argue that the
near-linear IRX-$M_\star$ relationship is a better proxy for
correcting observed UV luminosities to total star formation rates,
provided an accurate handle on $M_\star$ can be had, and also gives
clues as to the physical driver of the role of dust-obscured star
formation in high-redshift galaxies.
\end{abstract}

\begin{keywords}
dust, extinction -- galaxies: evolution, high-redshift, star formation, ISM -- cosmology: observations
\end{keywords}

%%%%%%%%%%%%%%%%%%%%%%%%%%%%%%%%%%%%%%%%%%%%%%%%%%

%%%%%%%%%%%%%%%%% BODY OF PAPER %%%%%%%%%%%%%%%%%%

\section{Introduction}
\label{sec:intro}

Understanding the evolution of the star formation rate density (SFRD)
with cosmic time has long been the cornerstone of extragalactic
astrophysics (e.g. \citealt{Madau_2014}). At $z>2$ most studies of the
evolution of the SFRD are based on samples of Lyman-break galaxies
(LBGs), due in part because of the efficiency of their selection
technique in deep broad band imaging surveys.

%where the photometry in three broad-band filters is used to
%detect the Lyman break -- the characteristic drop in the rest-frame UV
%spectrum, caused by the absorption by the intergalactic and
%interstellar medium at $\lambda<912\AA$. While at $z\sim 3$ the Lyman
%break is being detected, it is important to notice that at higher
%redshifts it is mainly the Lyman-alpha limit at 1216\AA\, that the LBG
%selection technique is sensitive to.

As a result, LBGs have been extensively studied and well-characterised
over the past two decades. They have stellar masses
$\sim$$10^{9-11}\,{\rm M_\odot}$ and star formation rates (SFRs)
$\sim$$10-100\,{\rm M_\odot\,y^{-1}}$, (e.g. \citealt{Madau_1996,
  Steidel_1996, Sawicki_1998, Shapley_2001, Giavalisco_2002,
  Blaizot_2004, Shapley_2005, Reddy_2006, Rigopoulou_2006, Verma_2007,
  Magdis_2008, Stark_2009, Chapman_2009, LoFaro_2009, Magdis_2010,
  Rigopoulou_2010, Pentericci_2010, Oteo_2013, Bian_2013}). LBGs are
therefore believed to be responsible for forming a substantial
fraction of massive local galaxies ($L>L^\ast$;
e.g. \citealt{Somerville_2001, Baugh_2005}), while those with the
highest SFRs ($>100\,{\rm M_\odot\,y^{-1}}$) could be the progenitors
of present-day ellipticals (e.g. \citealt{Verma_2007, Stark_2009,
  Reddy_2009}).

Naturally, given their selection, the most common tracer of LBGs' SFRs
has traditionally been through their rest-frame UV stellar continuum
emission (e.g. \citealt{Kennicutt_2012}). However, it is now well
known that about half of the starlight in the Universe is absorbed by
interstellar dust and re-emitted in the rest-frame far-infrared
(e.g. \citealt{Dole_2006}). It is therefore necessary to complement
UV-derived SFRs with far-infrared and sub-millimetre observations to
obtain a full census of star formation, with the latter providing the
most efficient probe of thermal dust emission out to high redshift
owing to the negative k-correction. Unfortunately, typical LBGs are
faint in the sub-millimetre, far below the confusion limit of most
single-dish sub-millimetre facilities and challenging even for
senstive interferometric facilities such as the Atacama Large
Millimeter/sub-millimeter Array (ALMA)\citep{Chapman_2000, Capak_2015,
  Bouwens_2016b, Koprowski_2016, Dunlop_2017, McLure_2017}. As a
result, representative samples of sub-millimeter-detected LBGs are not
available.

Without direct detection of the obscured star formation in individual
LBGs, empirical recipes are used to correct UV-derived SFRs to total
SFRs. The most common approach is to use the relationship between the
rest-frame UV slope, $\beta$, where $f_\lambda\propto\lambda^\beta$,
and the infrared excess, ${\rm IRX}\equiv L_{\rm IR}/L_{\rm
  UV} $\citep{Meurer_1999}. \citet{Overzier_2011} found that local
analogues of LBGs are consistent with the \citet{Meurer_1999}
relation, while at ($z\gtrsim 3$) \citet{Coppin_2015} and
\citet{Alvarez_2016} found LBGs to be lying above and below the local
relation respectively. Recently, \citet{McLure_2017} showed that the
IRX-$\beta$ relation for $z\sim 3$ galaxies is consistent with a
Calzetti-like attenuation law \citep{Calzetti_2000}, while
\citet{Reddy_2017} suggest that a flatter, Small Magellanic Cloud
(SMC)-like curve should be applied. In addition, a number of
individual direct detections for LBGs and infrared-selected galaxies,
have been found to exhibit a large scatter in the IRX-$\beta$ plane
\citep{Casey_2014b, Capak_2015, Scoville_2016, Koprowski_2016,
  Fudamoto_2017}. It remains unclear whether these inconsistencies are
due to intrinsic scatter in the IRX-$\beta$ relation or biases in the
selection and measurement technique. In either case, it is important
to re-evaluate the IRX-$\beta$ relation for high-$z$ LBGs if we wish
to use it for the accurate correction of UV-derived SFRs where direct
infrared detection is not available.

In this paper we make use of a large sample of LBGs at redshifts
$3<z<5$ in the UKIDSS/UDS field, stellar mass complete down to a limit
of ${\rm log}(M_\ast/{\rm M_\odot})\gtrsim 10.0$, to establish the
IRX--$\beta$ relation. We are able to determine the IR luminosities
for these galaxies through stacking of a deep JCMT SCUBA-2 850\,${\rm
  \mu m}$ map from the SCUBA-2 Cosmology Legacy Survey \citep{Geach_2017}, and 350--500$\mu$m SPIRE mapping from the {\it Herschel
  Space Observatory}. This paper expands on the work of
\citet{Coppin_2015}, with an improved SCUBA-2 map of UDS, now
approaching the SCUBA-2 confusion limit (with a 1$\sigma$ depth of
0.9\,mJy\,beam$^{-1}$). Section~\ref{sec:data} summarises the data
used and explains our LBG selection criteria. In
Section~\ref{sec:properties} we explain how the spectral energy
distribution (SED) fitting is performed and derive the basic physical
properties of galaxies in the sample. In Section~\ref{sec:disc} we
measure the IRX-$\beta$ relation for LBGs at $z=3$, $4$ and $5$ and
explain its physical origin, comparing our findings with other results
from the literature. We present our conclusions in
Section~\ref{sec:sum}.

Throughout, magnitudes are quoted in the AB system \citep{Oke_1983}
and we use the \citet{Chabrier_2003} stellar initial mass function
(IMF). We assume a cosmology with $\Omega_{\rm m} = 0.3$,
$\Omega_\Lambda = 0.7$ and $H_0 = 70$\,km\,s$^{-1}$\,Mpc$^{-1}$. We note that assuming the best-fit \citet{Planck_2016} cosmology yields $\simeq 2-2.5\%$ higher luminosity distances and hence $\simeq 4-5\%$ higher stellar masses and luminosities.

\section{Data}
\label{sec:data}

\subsection{Optical \& near-IR imaging}
\label{sec:oirdata}

Our sample is drawn from the deep $K$-band image of the UKIRT Infrared
Deep Sky Survey (UKIDSS; \citealt{Lawrence_2007}),
UDS\footnote{\url{http://www.nottingham.ac.uk/astronomy/UDS/}} data
release 8 (DR8), together with the available multi-wavelength
photometry. The UDS is the deepest of the five UKIDSS sub-surveys
(Almaini et al., in prep.), covering 0.77 deg$^2$ in the $J$, $H$ and
$K$ bands. The DR8 release achieves 5$\sigma$ point source depths of
24.9, 24.2 and 24.6\,mag, respectively. The parent catalogue was
extracted from the $K$-band image using SE{\sc xtractor}
\citep{Bertin_1996}. Two catalogues were constructed and merged: the
first was designed to extract point sources, while the second was
optimized to detect resolved galaxies (see \citealt{Hartley_2013} for
details). The UKIDSS UDS has also been imaged by the
Canada-France-Hawaii Telescope (CFHT) Megacam $U$-band (26.75\,mag),
Subaru Suprime-cam ($B = 27.6$, $V = 27.2$, $R = 27.0$, $i' = 27.0$,
and $z' = 26.0$; \citealt{Furusawa_2008}) and the {\it Spitzer}
Infrared Array Camera (IRAC; \citealt{Fazio_2004}, $[3.6{\rm \mu
    m}]=24.2$ and $[4.5{\rm \mu m}]=24.0$), as a part of the UDS {\it
  Spitzer} Legacy Program (SpUDS; PI:\,Dunlop). To remove obvious
active galactic nuclei (AGN), X-ray \citep{Ueda_2008} and radio
\citep{Simpson_2006} data were used. The total coincident area of
these data sets is 0.62 deg$^2$. All images were astrometrically
aligned and multi-band photometry extracted in 3-arcsec diameter
apertures at the positions of $K$-band detections (see
\citealt{Simpson_2012} for details), including point spread function
corrections where appropriate \citep{Hartley_2013}.

\subsubsection{LBG selection}
\label{sec:lbg}

The LBG selection technique relies on the fact that photons with
energies higher than the rest-frame $1216\AA$ are almost entirely
absorbed by the neutral gas around the star-forming regions in the
galaxy. This results in the characteristic break which can be easily
identified with broadband colours. This technique is primarily used to
identify galaxies at $z\approx 3$ using {\it UGR}, or {\it BVR},
filters \citep{Steidel_1996}, but can be easily extended to higher
redshifts by simply shifting the colour space to longer wavelengths,
as described by \citet{Ouchi_2004}. In this work we use the following
selections for LBGs at $z\approx 3$ (Equation~\ref{eq:lbgs1}),
$z\approx 4$ (Equation~\ref{eq:lbgs2}) and $z\approx 5$
(Equations~\ref{eq:lbgs3} and \ref{eq:lbgs4}):

\begin{center}
\begin{equation}\label{eq:lbgs1}
\begin{array}{ll}
R<27, & (U -V)>1.2, \\
-1.0<(V -R)<0.6, &
(U-V)>3.8 (V -R) + 1.2;\\	
\end{array}
\end{equation}
\end{center}

\begin{center}
\begin{equation}\label{eq:lbgs2}
\begin{array}{ll}
i'<27, & (B -R)>1.2, \\
(R -i')<0.7, & 
(B -R)>1.6(R -i') + 1.9;\\	
\end{array}
\end{equation}
\end{center}

\begin{center}
\begin{equation}\label{eq:lbgs3}
\begin{array}{ll}
z'<26, & (V-i')>1.2, \\
(i' -z')<0.7, & 
(V -i')>1.8(i'-z') + 2.3;\\	
\end{array}
\end{equation}
\end{center}

\begin{center}
\begin{equation}\label{eq:lbgs4}
\begin{array}{ll}
z'<26, & (R -i')>1.2, \\
(i' -z')<0.7, & 
(R -i')>(i' -z') + 1.0,\\	
\end{array}
\end{equation}
\end{center}

\noindent where $z\approx 5$ LBGs are identified using either
Equation~\ref{eq:lbgs3} or \ref{eq:lbgs4} in order to maximise our
yield (see \citealt{Ouchi_2004}). Note that, since the parent optical
catalogue is selected at $K$-band ($K<24.6$), our resulting LBG sample
is mass complete to a limit of ${\rm log}(M_\ast/{\rm M_\odot})\gtrsim
10.0$.

Photometric redshifts are determined for each source in our parent
catalogue using 11 photometry bands ($UBVRi'z'JHK[3.6][4.5]$), as
described in \citet{Hartley_2013} and \citet{Mortlock_2013}, using the
{\sc eazy} template-fitting code. Six SED templates were used
\citep{Brammer_2008}, with the bluest template having a SMC-like
extinction added. The accuracy of the photometric redshifts is
assessed by comparing with the available spectroscopic data, as
discribed in \citet{Hartley_2013}, with the average $|z_{\rm
  phot}-z_{\rm spec}|/(1+z_{\rm spec})=0.031$.

\begin{figure*}
\begin{center}
\includegraphics[scale=0.7]{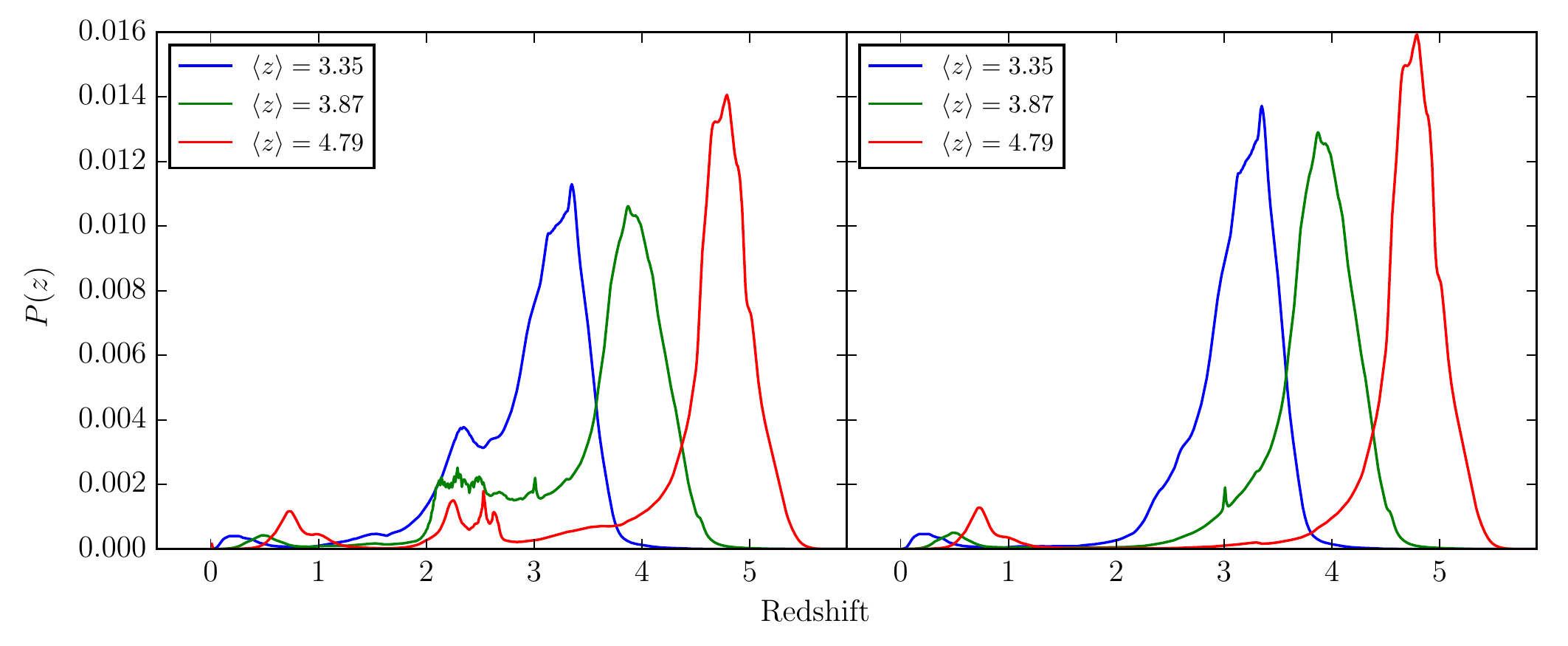}
\end{center}
\caption{Redshift probability distributions with the corresponding
  most-probable redshifts shown in the legend. {\bf Left:} Redshift
  probability distributions for the LBG selection criteria from
  Equations~\ref{eq:lbgs1}-\ref{eq:lbgs4}, with the additional
  constraint of $z>2$ in place. It can be seen that the resulting
  most-probable redshifts are close to the target values of 3, 4 and
  5. However, the distributions show a low-redshift peak at $z\simeq
  2.5$, this being the result of a number of contaminating galaxies
  being included using our selection criteria. {\bf Right:} Since the
  $z\simeq 2.5$ sources from the left panel will contaminate the
  inferred values of the stellar masses, as well as UV slopes, we
  decided to introduce an additional selection criteria, where we
  force the redshifts to be $>2.5$, $>3$ and $>4$ for the $z\sim3$,
  $z\sim4$ and $z\sim5$ samples, respectively. This panel shows the
  resulting redshift probability distributions. Note that the
  low-redshift peaks at $z\sim0.5$ do not result from a number of
  sources being found at low redshifts, but rather from a small number
  of individual probability distributions being double-peaked (with
  the low-$z$ solution having lower probability).}
\label{fig:z}
\end{figure*}

To help eliminate low redshift interlopers in the LBG selections, we
initially enforce the minimum best-fit (i.e.\ peak of the redshift
probability density distribution) redshift to be $z=2$. In the left
panel of Figure~\ref{fig:z} the normalised sum of the redshift
probability distributions are shown for each redshift selection,
indicating peaks at 3.35, 3.87 and 4.79. Thus, the selection criteria
used here selects galaxies at redshifts consistent with the target
values. However, all three distributions show a minor peak at
$z\approx2.5$, indicating contamination still present in the
selection. To remedy this situation, we further enforce the maximum
likelihood redshifts ($z_{\rm best}$) to be $z>2.5$, $z>3$ and $z>4$
for the $z\approx 3$, $z\approx 4$ and $z\approx 5$ samples,
respectively. This results in much `cleaner' redshift probability
distributions containing 3419, 699 and 60 sources at mean redshifts of
3.35, 3.87 and 4.79, respectively. For the reasons explained in the next Section, we exclude 36 LBGs that are
directly detected in the rest-frame FIR in high-resolution ALMA
follow-up of all S2CLS-detected 850\,$\mu$m sources in the UDS field
(PI: Smail). A detailed study of these ALMA-detected LBGs will be
presented in Koprowski et al. (in preparation).

%For reasons explained in the next
%section, we have also excluded 36 LBGs that are directly detected in
%the rest-frame FIR in high-resolution ALMA follow-up of all
%S2CLS-detected 850$\mu$m sources in the UDS field (PI: Smail). A
%detailed study of these ALMA-detected LBGs will be presented in
%Koprowski et al. (in preparation).

%Comparing to \citet{Coppin_2015}, due to the refined redshift
%selection, the sample was reduced by 782 (19\%) at $z=3.35$, 170
%(20\%) at $z=3.87$ and 8 (12\%) at $z=4.79$. In
%Table\,\ref{tab:fluxes} two values for the average 850\,${\rm \mu m}$
%flux densities at a given redshift are given: with and without the 36
%submm-detected LBGs in the sample. We use the latter in the remaining
%analysis of this paper, noting that the stacked values are
%approximately 40\% lower than those measured in
%\citet{Coppin_2015}. 

\subsection{IR \& sub-mm imaging}
\label{sec:ir}

\subsubsection{Spitzer MIPS \& Herschel SPIRE data}
\label{sec:spire}

\begin{table*}
\begin{footnotesize}
\begin{center}
\caption{Stacked IR-submm photometry for LBGs. The columns show the most-probable redshift in each bin, the number of selected LBGs and the stacked photometry in the {\it Herschel} SPIRE and JCMT SCUBA-2 850\,${\rm \mu m}$ bands, with $1\sigma$ errors and detection significance in brackets.}
\label{tab:fluxes}
\setlength{\tabcolsep}{2 mm} 
\begin{tabular}{ccccccc}
\hline
\hline
$\langle z_{\rm phot}\rangle$ & $N$ & $S_{250}$ & $S_{350}$                 & $S_{500}$                      & $S_{850}^{a}$                           & $S_{850}^{b}$                 \\
                              &     & /\,mJy    & /\,mJy                    & /\,mJy                         & /\,mJy                                  & /\,mJy                        \\
\hline
3.35 & 3419 & 0.534$\pm$0.076 (7.0$\sigma$) & 0.668$\pm$0.077 (8.7$\sigma$) & 0.360$\pm$0.085 (5.5$\sigma$) & 0.181$\pm$0.015 (12.1$\sigma$)           & 0.128$\pm$0.015 (8.5$\sigma$) \\
3.87 & 699  & 0.533$\pm$0.162 (3.3$\sigma$) & 0.957$\pm$0.171 (5.6$\sigma$) & 0.966$\pm$0.185 (5.5$\sigma$) & 0.369$\pm$0.033 (11.2$\sigma$)           & 0.261$\pm$0.033 (7.9$\sigma$) \\
4.79 & 60   & 0.294$\pm$0.556 (0.5$\sigma$) & 0.485$\pm$0.573 (0.8$\sigma$) & 0.517$\pm$0.618 (0.9$\sigma$) & 0.417$\pm$0.110 (3.8$\sigma$)\phantom{1} & 0.335$\pm$0.110 (3.0$\sigma$)  \\
\hline
\end{tabular}
\end{center}
\begin{flushleft}
$^a$ Mean weighted stack with the ALMA-detected LBGs present.\\
$^b$ Mean weighted stack with the ALMA-detected LBGs removed from the 850\,${\rm \mu m}$ map. We use this column for SED fitting.\\
\end{flushleft}
\end{footnotesize}
\end{table*}

\begin{figure*}
\begin{center}
\includegraphics[scale=0.31, trim=0cm 0cm 0cm 0cm]{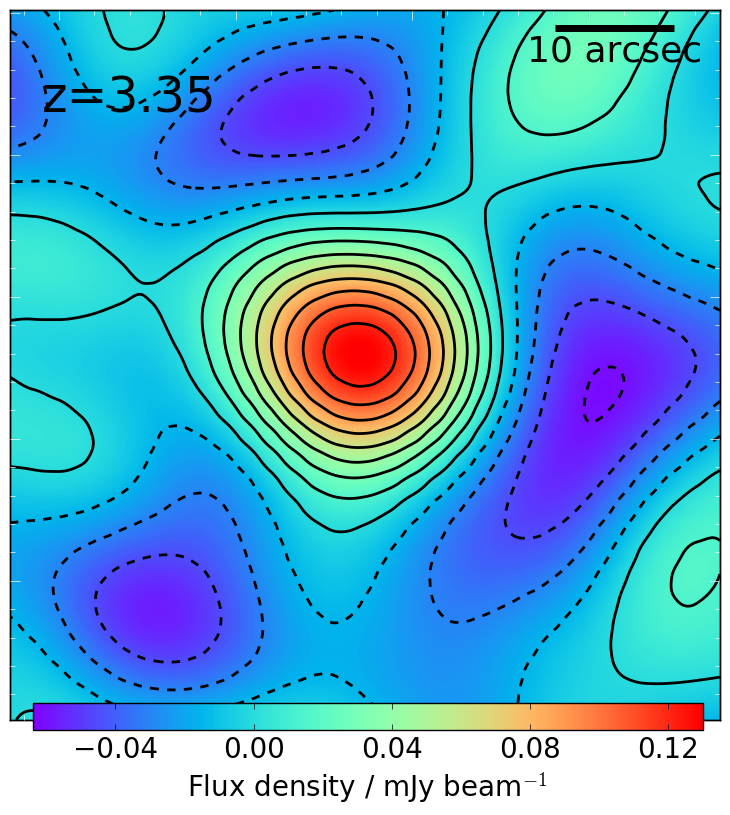}\includegraphics[scale=0.31, trim=0cm 0cm 0cm 0cm]{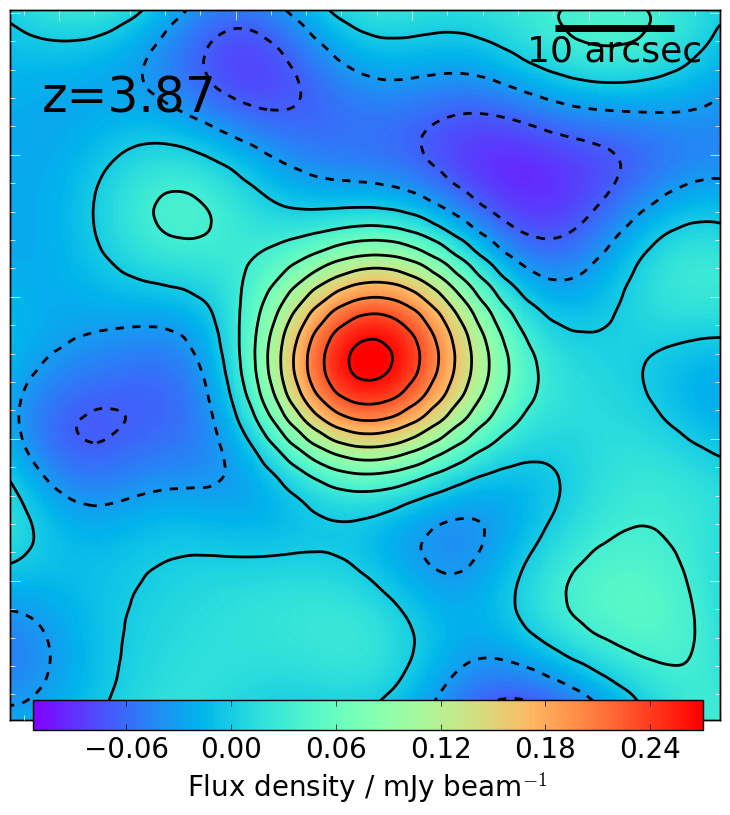}\includegraphics[scale=0.31, trim=0cm 0cm 0cm 0cm]{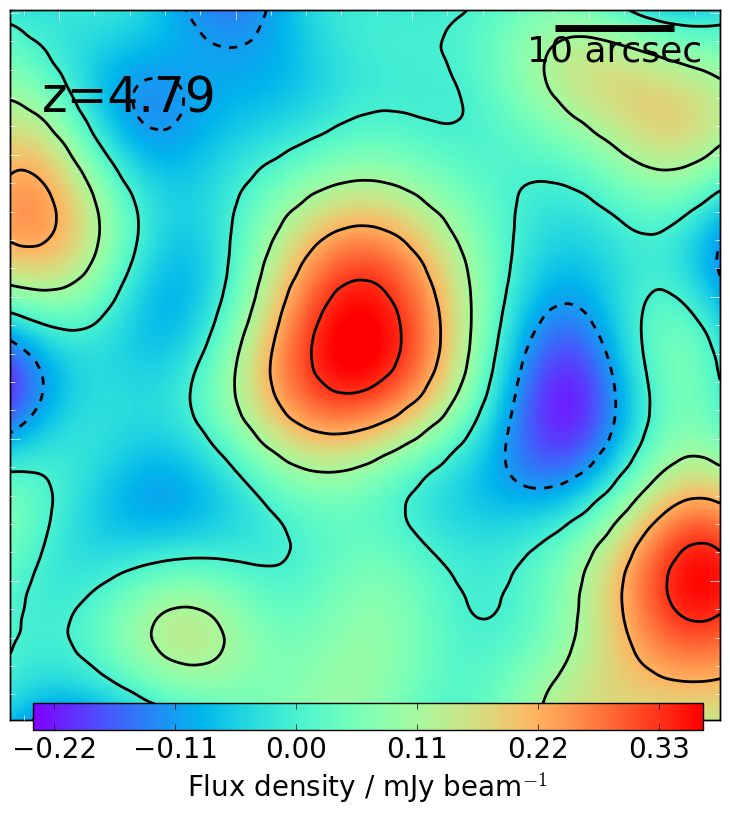}
\end{center}
\caption{60\,arcsec $\times$ 60\,arcsec stamps of the stacked 850\,${\rm \mu m}$ flux densities in the SCUBA-2 maps for each redshift bin, centred on the LBG positions. The solid contours represent significance levels of $0\sigma$, $1\sigma$, $2\sigma$,\dots, while the dashed lines show negative deviations. We find significant detections in each redshift bin, with amplitudes $8.5\sigma$, $7.9\sigma$ and $3.0\sigma$ at $z=3.36$, $z=3.87$ and $z=4.79$, respectively (see Table~\ref{tab:fluxes}).}
\label{fig:stamps}
\end{figure*}

We utilise mid-IR imaging from the Multiband Imaging Photometer for
{\it Spitzer} instrument (MIPS; \citealt{Rieke_2004}) at 24\,${\rm \mu
  m}$ from the {\it Spitzer} Public Legacy Survey of the UKIDSS Ultra
Deep Survey (SpUDS; PI: J. Dunlop), as described in
\citet{Caputi_2011}, and sub-millimetre imaging from {\it Herschel}
\citep{Pilbratt_2010}, as provided by the public release of the HerMES
\citep{Oliver_2012} survey undertaken with the SPIRE
\citep{Griffin_2010} instrument, at 250, 350 and 500\,${\rm \mu
  m}$. The Level 2 data products from the {\it Herschel} European
Space Agency (ESA) archive were retrieved, aligned and co-added to
produce maps. The SPIRE maps are confused, and so we apply a
de-blending procedure following \citet{Swinbank_2014} using sources in
the 24\,$\mu$m catalogue as priors for the positions of sources
contributing to the SPIRE map. In brief, the optimal sky model is
found assuming 24\,$\mu$m sources contribute to SPIRE sources detected
at $>$2$\sigma$ at 250\,$\mu$m and 350\,$\mu$m by minimising the residual
flux density between a (PSF-convolved) sky model and the data. The
best fitting 24\,$\mu$m sources are then subtracted from the SPIRE maps,
excluding those actually associated with LBGs in our samples. This
minimises the contribution to the stacked flux at the position of LBGs
from confused SPIRE sources. 

In addition, we have also decided to remove all the SPIRE-detected sources. This is motivated by the fact that the most IR-bright galaxies will lie significantly above the IRX-$\beta$ relation and hence a simple dust-screen model used in the derivation of the functional form of the relation \citep{Meurer_1999} cannot be applied to them. \citet{Casey_2014b} found that virtually all the galaxies with the IR luminosity ${\rm log}(L_{\rm IR}/{\rm L_\odot})\gtrsim 11.5$ have their IRX values significantly elevated above the local IRX-$\beta$ relation. With our average SEDs found in Section\,\ref{sec:sed}, we find that all the LBGs detected at SPIRE bands will satisfy the above criterion and hence the removal of all the SPIRE-detected sources is well motivated. 

We have therefore excluded all the SPIRE galaxies from the maps, but we note that leaving the sources associated with our LBGs in increases the resulting values of IRX only by $\sim 20\%$. The resulting stacked SPIRE flux densities with errors are given in Table~\ref{tab:fluxes}.

\subsubsection{JCMT SCUBA-2 data}
\label{sec:scuba}

\begin{figure}
\begin{center}
\includegraphics[scale=0.7]{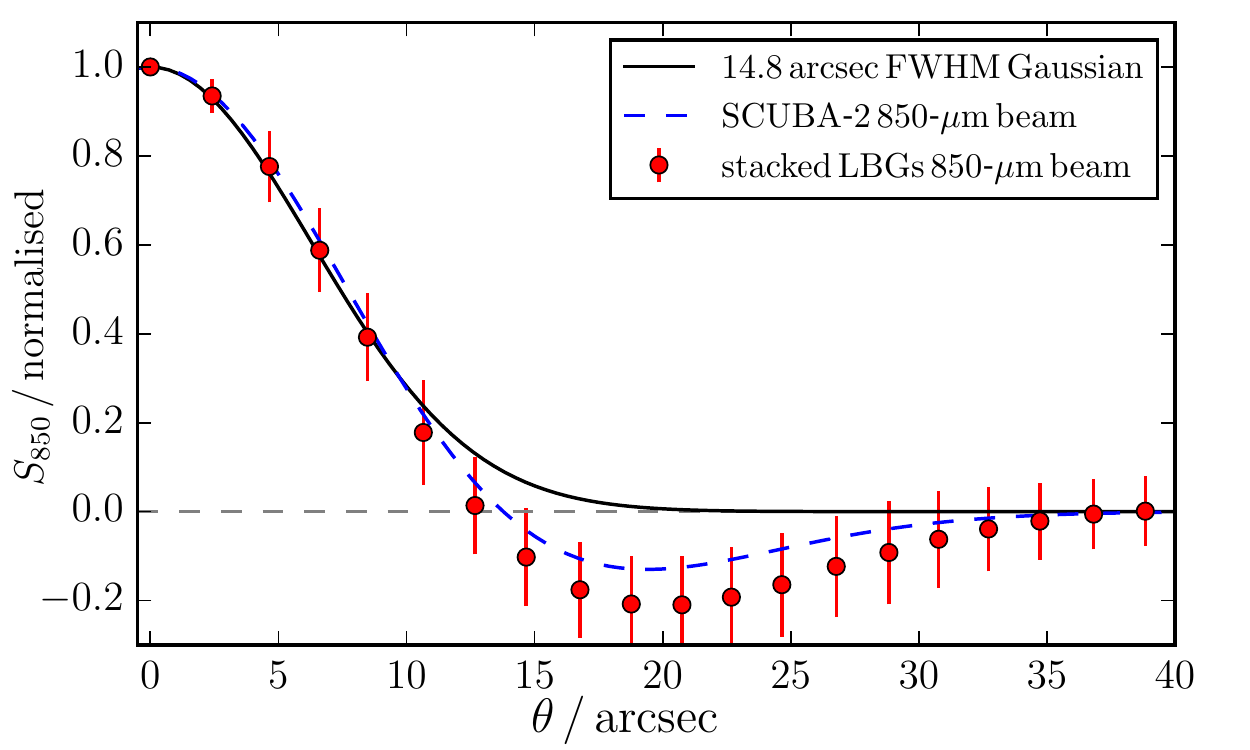}
\end{center}
\caption{Light profile of the $z\approx 3$ 850-${\rm \mu m}$ stack
from the left panel of Figure~\ref{fig:stamps} (red points),
compared with the SCUBA-2 850-${\rm \mu m}$ beam (dashed blue line)
of \citet{Geach_2017} and a 14.8\,arcsec FWHM Gaussian (black solid
curve). The stacked profile is consistent with the beam and
therefore any contribution from clustering of associated sources
must be on scales below 15$''$ if present at all.}
\label{fig:beams}
\end{figure}

\begin{figure*}
\begin{center}
\includegraphics[scale=0.8]{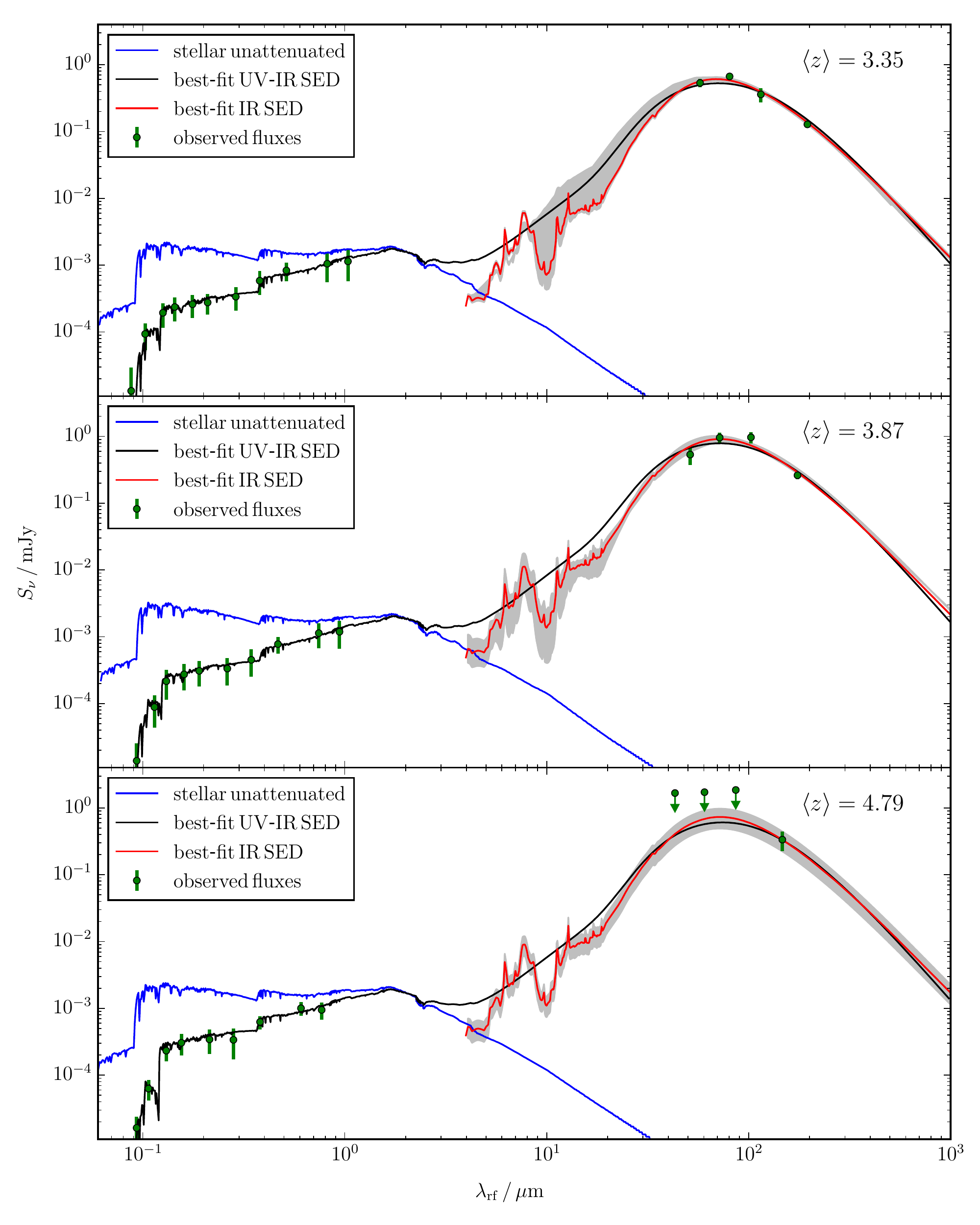}
\end{center}
\caption{Best-fit SEDs at rest-frame wavelengths for the stacked results in each redshift bin. The IR photometry comes from stacking LBGs in {\it Herschel} SPIRE and JCMT SCUBA-2 850-${\rm \mu m}$ bands, while the rest-frame UV-NIR photometry points are median values from all the LBGs in a given redshift bin, with the errors being median absolute deviations. The red curves (used in the calculations of $L_{\rm IR}$) are best-fit empirical IR SEDs of \citet{Swinbank_2014} found using a $\chi^2$ minimisation method. In addition, we plot in black the best-fit rest-frame UV-mm SEDs found using {\sc cigale}, where \citet{Bruzual_2003} stellar population templates, \citet{Chabrier_2003} IMF, \citet{Calzetti_2000} dust attenuation law and the thermal dust emission model of \citet{Casey_2012c} were adopted (see Section~\ref{sec:sed} for details). Since {\sc cigale} uses energy balance, we can find the amount of rest-frame UV dust reddening based on the IR luminosity and therefore estimate the unattenuated stellar emission SED, which we show here in blue.}
\label{fig:seds}
\end{figure*}

We use the final, near-confusion-limited 850$\,\mu$m map of the UDS
from the SCUBA-2 Cosmology Legacy Survey (S2CLS). Full details of the
observations and data reduction are given in \citet{Geach_2017}, but
the map spans 1\,degree centred on the UDS and reaches a uniform depth
of 0.9\,mJy\,beam$^{-1}$. Note that this final map is a factor of 2
deeper than the map used in \citet{Coppin_2015}. We subtracted all
SCUBA-2 sources with the signal-to-noise ratio (SNR) of $>3.5$ from
the 850-${\rm \mu m}$ maps following the same reasoning as the
de-blending procedure used in the SPIRE maps.

We evaluate the stacked flux density of LBGs in SCUBA-2 and SPIRE maps
following an inverse variance weighting:

\begin{equation}\label{eq:stack} \langle S_\nu\rangle = \frac{\Sigma^N_i S_{\nu,i}/\sigma_i^2}{\Sigma^N_i 1/\sigma_i^2}, \end{equation}

\noindent where $S_{\nu,i}$ is the flux density of the $i$th LBG and
$\sigma_i$ is the $1\sigma$ instrumental noise at the same position.
We assume that confusion noise is constant over the field. The error
on the mean is then found from

\begin{equation}\label{eq:stackerr} \sigma^2(\langle S_\nu\rangle)=\frac{1}{\Sigma^N_i 1/\sigma_i^2}. \end{equation}

\noindent We show thumbnail images in Figure~\ref{fig:stamps}
presenting the inverse-variance weighted 850$\mu$m stacks at the
positions of LBGs in the three redshift bins, and list the average
flux densities in Table~\ref{tab:fluxes}. We find $8.5\sigma$,
$7.9\sigma$ and $3.0\sigma$ detections in the $z\approx 3$, $z\approx
4$ and $z\approx 5$ redshift bins,
respectively. Table~\ref{tab:fluxes} also gives the stacked SPIRE flux
densities where we measure significant stacked detections in all but
the $z\approx5$ bin. 

Figure~\ref{fig:beams} shows the average radial profile of the
$z\approx3$ stack compared to the SCUBA-2 beam (which differs slightly
from a pure Gaussian). The stacked profile is indistinguishable from
the shape of the beam and therefore any clustering of sources
associated with the LBGs that also contribute to the 850$\mu$m flux
density \citep{Chary_2010, Kurczynski_2010, Serjeant_2010} are on
scales unresolved by SCUBA-2; i.e. below approximately 15$''$ or
120\,kpc. We ignore this potential contribution in the following
analysis and consider the average submillimetre emission as coming
from the LBG itself.

\section{Results}
\label{sec:properties}

\subsection{SED fitting}
\label{sec:sed}

\begin{figure*}
\begin{center}
\includegraphics[scale=0.7]{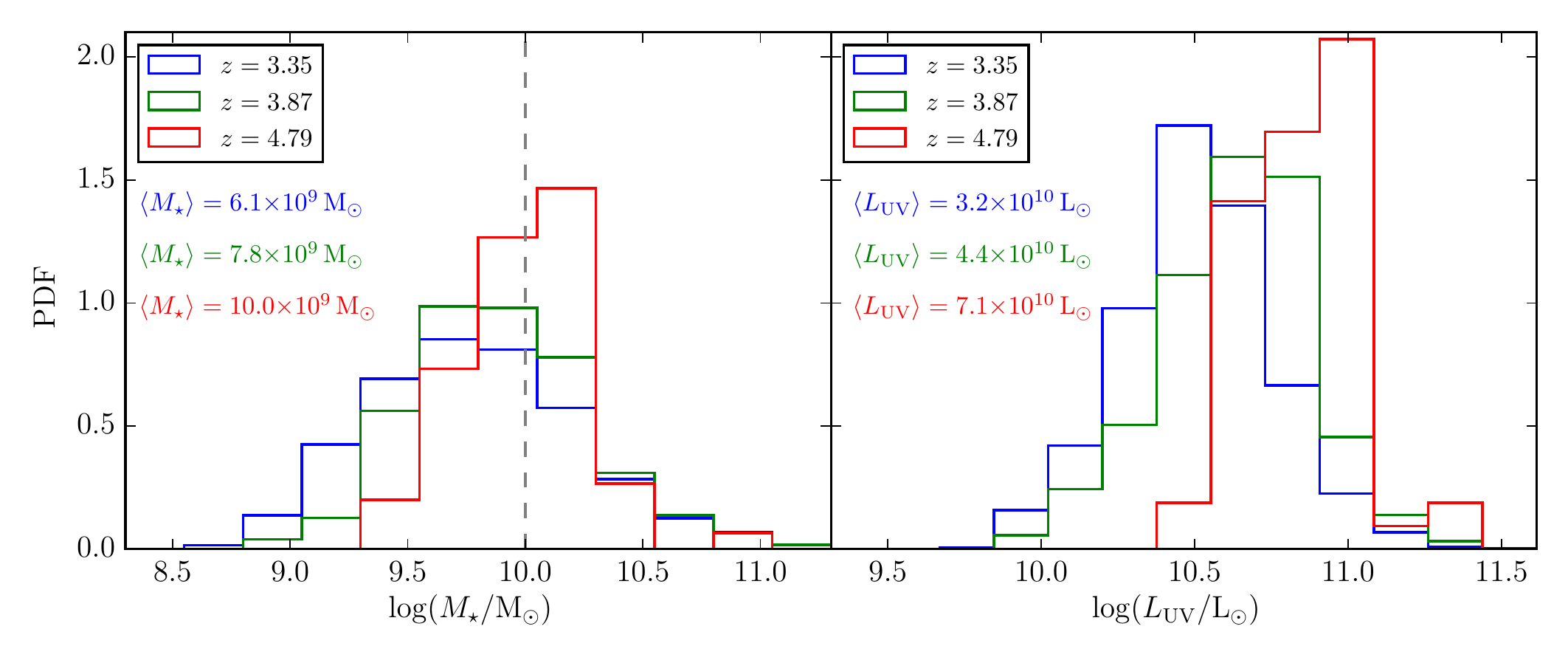}
\end{center}
\caption{{\bf Left:} Histogram of stellar masses for each redshift bin found from the best-fit rest-frame UV-NIR SEDs (see Section~\ref{sec:props} for details). It can be seen that the masses increase with redshift, which is a consequence of the selection limit for our parent catalogue of $K<24.6$. Since we can treat $K$-band as a rough proxy for the stellar mass (see Figure\,\ref{fig:massK}), we expect the average mass to increase with redshift due to the positive $K$-correction in this band. The grey dashed line marks the mass limit down to which our LBG sample is complete. This is a consequence of our parent optical catalogue being selected at $K$-band (see also Figure\,\ref{fig:massK}). {\bf Right:} Histogram of the UV luminosities for each redshift bin in this work. As in the case of the stellar masses, the UV luminosities tend to increase with redshift. Again, this is caused by the depth of our parent catalogue in the rest-frame UV bands.}
\label{fig:hist}
\end{figure*}

\begin{figure}
\begin{center}
\includegraphics[scale=0.7]{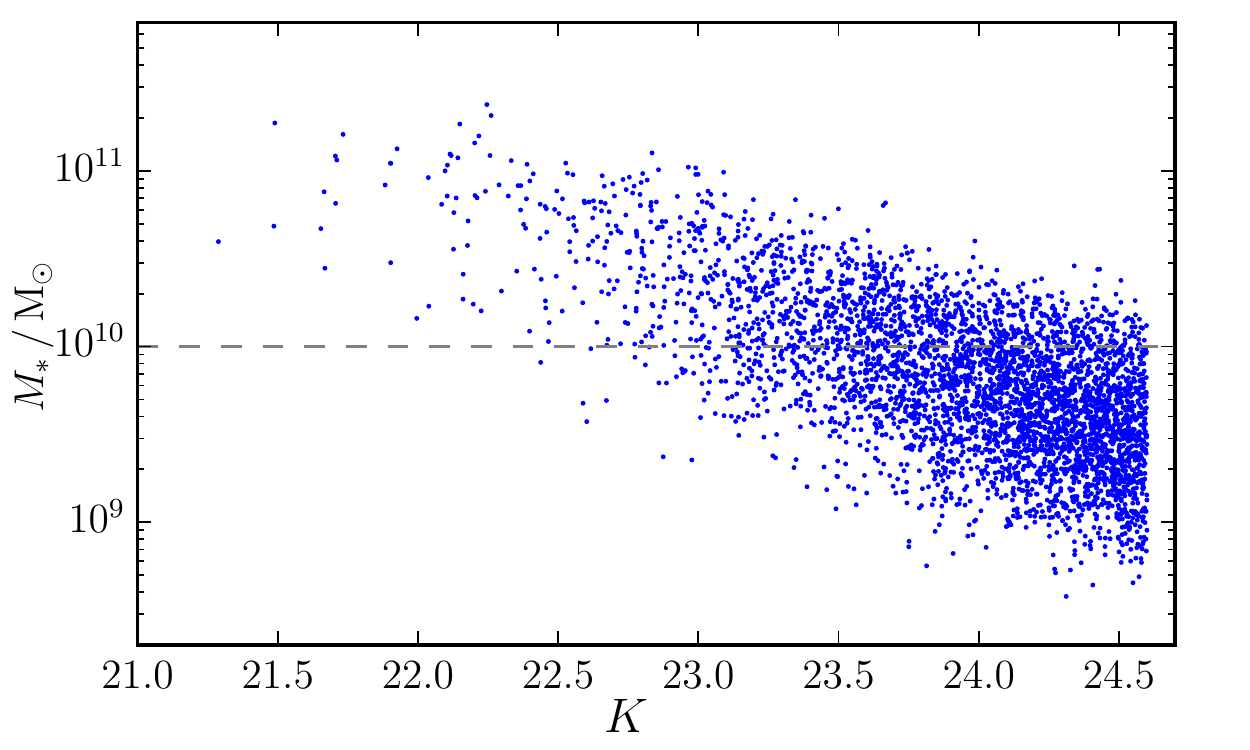}
\end{center}
\caption{Stellar mass as a function of the $K$-band magnitude for the whole LBG sample used in this work. It can be seen that the $K$-band can be treated as a very rough proxy of the stellar mass. Applying a $K$-band cut of 24.6 (Section\,\ref{sec:oirdata}) at all 3 reshift bins causes slightly higher stellar masses to be selected at higher redshift bins, due to the positive $K$-correction (see Figure\,\ref{fig:hist} and Table\,\ref{tab:props}). The dashed grey line marks the mass completeness limit, being the consequence of our parent optical catalogue being selected at $K$-band.}
\label{fig:massK}
\end{figure}

To fit the stacked flux densities we use 185 SED templates compiled by
\citet{Swinbank_2014}. These include local galaxy templates from
\citet{Chary_2001}, \citet{Rieke_2009} and \citet{Draine_2007}, as
well as high-redshift starburst galaxies from \citet{Ivison_2010} and
\citet{Carilli_2011}, with a range of dust temperatures spanning
19-60\,K. With redshifts fixed at the peak values from
Table~\ref{tab:fluxes} we find the best-fitting SEDs using a standard
$\chi^2$ minimisation approach. At $z=4.79$ only the 850-${\rm \mu m}$
stacked flux density was detected at $>$$3\sigma$ and so here we adopt
our $z=3.87$ best-fitting SED redshifted to $z=4.79$ and normalised to
the 850$\mu$m flux. The fits have a consistent temperature of $T_{\rm d}\approx$40\,K.

We also determine the best-fitting rest-frame UV-to-mm model SEDs,
where the UV-through-NIR photometry and uncertainties are medians and
median absolute deviations for all LBGs in the redshift bin. We use
the `energe balance' code {\sc
  cigale}\footnote{\url{http://cigale.lam.fr/}}
\citep{Noll_2009,Serra_2011}, adopting the \citet{Bruzual_2003}
stellar population templates with a double-burst, exponentially
declining star formation history (SFH) in which the dependence of star
formation rate on time is

\begin{equation}\label{eq:sfr} \Psi(t) \propto {\rm exp}(-t_1/\tau_1)+f_{\rm m}{\rm exp}(-t_2/\tau_2), \end{equation}

\noindent with $\tau_1$, $\tau_2$ and the mass fraction of the late
burst population, $f_{\rm m}$, being free parameters. This allows a large
variation of in SFH, allowing for both single-burst and double-burst,
instantaneous, exponentially declining and continuous histories. Dust
attenuation is implemented using \citet{Calzetti_2000} prescriptions
and thermal dust emission uses the model of \citet{Casey_2012c}, where
the mid-infrared power-law slope and dust emissivity index are fixed
at 2.6 and 1.6, respectively, while the temperature is allowed to vary
between 20 and 80\,K\footnote{Note, that the dust temperature in the
  \citet{Casey_2012c} models is an effective temperature, which is
  significantly higher than the temperature corresponding to the peak
  of the thermal infrared emission (see Figure 2 in
  \citealt{Casey_2012c}).}. The best-fit SEDs are plotted in
Figure~\ref{fig:seds} as black curves. Since {\sc cigale} uses energy
balance, the corresponding unattenuated stellar emission can be
estimated, which we show in Figure~\ref{fig:seds} as blue curves.

\subsection{UV \& IR luminosities and stellar masses}
\label{sec:props}

The {\sc cigale} fits described above are used to estimate the average
stellar mass of each sample. As noted by \citet{Dunlop_2011}, the use
of a multi-component SFH generally leads to more accurate values of
stellar mass than the use of a single SFH. This is due to the fact
that in a single burst scenario the entire stellar population must be
young in order to reproduce the UV emission, thus the less massive but
more abundant old stars are often not properly accounted for (see also
\citealt{Michalowski_2012a, Michalowski_2014}). The stellar mass
distributions and corresponding mean values for each redshift bin are
shown in Figure~\ref{fig:hist}, with the numbers summarised in
Table~\ref{tab:props}. The average stellar mass increases with
redshift, which is a simple consequence of the NIR selection limit for
our parent catalogue (Section~\ref{sec:oirdata}), as shown in
Figure\,\ref{fig:massK}. For the same reason, our $K$-band limited
sample is only complete to a stellar mass limit of ${\rm
  log}(M_\ast/{\rm M_\odot})\gtrsim 10.0$ (see
Figure\,\ref{fig:massK}).

The UV luminosity is defined here as $L_{\rm UV}\equiv
\nu_{1600}L_{1600}$, where the luminosity density at rest-frame
1600\AA, $L_{1600}$, is determined from the best-fitting SED. The
luminosity distributions are shown in the right panel of
Figure~\ref{fig:hist}, with the mean values summarised in
Table~\ref{tab:props}. Again, $L_{\rm UV}$ is increasing with
redshift, which, as in the case of the stellar mass, is a result of
the fixed optical flux limits in the LBG selection. While the
difference between $z=3.35$ and $z=3.87$ is small ($R<27$ and $i'<27$
from Equations~\ref{eq:lbgs1} and \ref{eq:lbgs2}, respectively), the
UV luminosity for $z=4.79$ is significantly higher because the
corresponding rest-frame UV imaging is shallower ($z'<26$,
Equations~\ref{eq:lbgs3} and \ref{eq:lbgs4}).

Finally, total IR luminosities are determined by integrating under the
best-fitting IR SED between rest-frame 8 and 1000\,${\rm \mu m}$
(Table~\ref{tab:props}). Again, average $L_{\rm IR}$ increases with
redshift, which is most likely linked to the increase in stellar mass,
rather than a real evolutionary trend.

\begin{table}
\begin{footnotesize}
\begin{center}
\caption{Physical properties for LBGs. The stellar masses and UV luminosities are mean values in each bin (see Figure~\ref{fig:hist}), with the errors being standard deviations rather than the errors on the mean (gives indication of the scatter). The IR luminosities are found by integrating the best-fit empirical IR templates (red curves in Figure~\ref{fig:seds}) between 8 and 1000\,${\rm \mu m}$.}
\label{tab:props}
\setlength{\tabcolsep}{2 mm} 
\begin{tabular}{cccc}
\hline
\hline
$\langle z\rangle$ & ${\rm log}(M_\ast/{\rm M}_\odot)$ & ${\rm log}(L_{\rm UV}/{\rm L}_\odot)$ & ${\rm log}(L_{\rm IR}/{\rm L}_\odot)$ \\
\hline
3.35                          & \phantom{1}$9.78\pm 0.45$               & $10.51\pm 0.25$                           & $11.36^{+0.14}_{-0.03}$                   \\
3.87                          & \phantom{1}$9.89\pm 0.38$               & $10.65\pm 0.27$                           & $11.64^{+0.05}_{-0.05}$                   \\
4.79                          & $10.00\pm 0.27$                         & $10.85\pm 0.17$                           & $11.69^{+0.12}_{-0.17}$                   \\
\hline
\end{tabular}
\end{center}
\end{footnotesize}
\end{table}

\begin{table}
\begin{footnotesize}
\begin{center}
\caption{Stacking results for our LBG sample. In each $\beta$ bin, the value of the UV slope is the unweighted average and the error bars correspond to the width of a given bin. A total number of LBGs in each $\beta$ bin are given and the IRX values and errors are calculated using Equations\,\ref{eq:stack} and \ref{eq:stackerr}, respectively. The stellar mass in each stellar mass bin is the mean with the error being the standard error on the mean. The lowest-mass bin upper limit is the only mass incomplete data point (see Figure\,\ref{fig:massK}).}
\label{tab:values}
\setlength{\tabcolsep}{4 mm} 
\begin{tabular}{lcc}
\hline
\hline
Sample                         & N    & $\langle {\rm IRX}\rangle$            \\
\hline
$\bmath{\beta\,{\rm bins}:}$   &      &                                       \\
\hspace{0.3cm} $z=3.35$        &      &                                       \\
\hspace{0.6cm} $\beta = -2.00$ & 1523 &            $<2.62$\phantom{1}         \\
\hspace{0.6cm} $\beta = -1.63$ & 1420 & \phantom{1}$6.42\pm 1.19$\phantom{1}  \\
\hspace{0.6cm} $\beta = -1.09$ & 312  &            $17.75\pm 4.21$\phantom{1} \\
\hspace{0.6cm} $\beta = -0.58$ & 111  &            $57.21\pm 10.04$           \\
\hspace{0.6cm} $\beta = -0.03$ & 33   &            $77.46\pm 21.56$           \\
\hspace{0.3cm} $z=3.87$        &      &                                       \\
\hspace{0.6cm} $\beta = -1.93$ & 228  &            $<3.92$\phantom{1}         \\
\hspace{0.6cm} $\beta = -1.57$ & 351  & \phantom{1}$7.76\pm 1.62$\phantom{1}  \\
\hspace{0.6cm} $\beta = -1.03$ & 81   &            $21.49\pm 5.14$\phantom{1} \\
\hspace{0.6cm} $\beta = -0.48$ & 27   &            $<44.92$                   \\
\hspace{0.3cm} $z=4.79$        &      &                                       \\
\hspace{0.6cm} $\beta = -1.74$ & 50   &            $<8.36$\phantom{1}         \\
\hspace{0.6cm} $\beta = -1.11$ & 7    &            $19.78\pm 6.10$\phantom{1} \\
$\bmath{M_\ast\,{\rm bins}:}$  &      &                                       \\
\hspace{0.3cm} ${\rm log}(M_\ast/{\rm M_\odot}) = \phantom{1}9.47\pm 0.13$  & 1339 &            $<2.90$\phantom{1}          \\
\hspace{0.3cm} ${\rm log}(M_\ast/{\rm M_\odot}) = \phantom{1}9.90\pm 0.13$  & 1633 & \phantom{1}$6.48\pm 1.13$\phantom{1}   \\
\hspace{0.3cm} ${\rm log}(M_\ast/{\rm M_\odot}) =           10.33\pm 0.13$  & 715  &            $23.08\pm 2.39$\phantom{1}  \\
\hspace{0.3cm} ${\rm log}(M_\ast/{\rm M_\odot}) =           10.81\pm 0.13$  & 157  &            $38.04\pm 6.00$\phantom{1}  \\
\hline
\end{tabular}
\end{center}
\end{footnotesize}
\end{table}

\section{Analysis \& Discussion}
\label{sec:disc}

\subsection{IRX-$\beta$ relation}
\label{sec:irx}

\subsubsection{UV slopes}
\label{sec:slopes}

Several different techniques have been used in the literature to
measure the UV slope, $\beta$ (see \citealt{Rogers_2013} for a
review). The original work of \citet{Meurer_1999} fitted a simple
power-law to the ten continuum bands listed by \citet{Calzetti_1994}
in the rest-frame range of $\sim$1250--2500\AA. In most cases,
however, only a few bands are available in that range, introducing
uncertainty on $\beta$. In addition, the possible existence of the
2175\AA\, feature in the dust attenuation curve can potentially impact
the inferred values of the photometry-based UV slopes, driving up
scatter in $\beta$. As shown by \citet{McLure_2017} and explained
further below, this scatter is significant enough to cause a bias that
serves to flatten the IRX-$\beta$ relation. To try to minimise these
effects, we measure $\beta$ by fitting a power-law to the best-fitting
SED over a rest-frame range of 1250-2500\AA, rather than the
photometry directly.

%Since the
%unattenuated stellar emission SED is very well described by a simple
%power law between $\sim$1200 and 2600\AA, any attenuation curve
%without the 2175\AA\, feature will result in the observed SEDs also
%being simple power-laws. Because the width of the possible UV bump
%($\sim 350$\AA) is significantly narrower than that range, the UV
%slopes based on the best-fit SEDs will be relatively insensitive to
%its presence.

\begin{figure*}
\begin{center}
\includegraphics[scale=0.8]{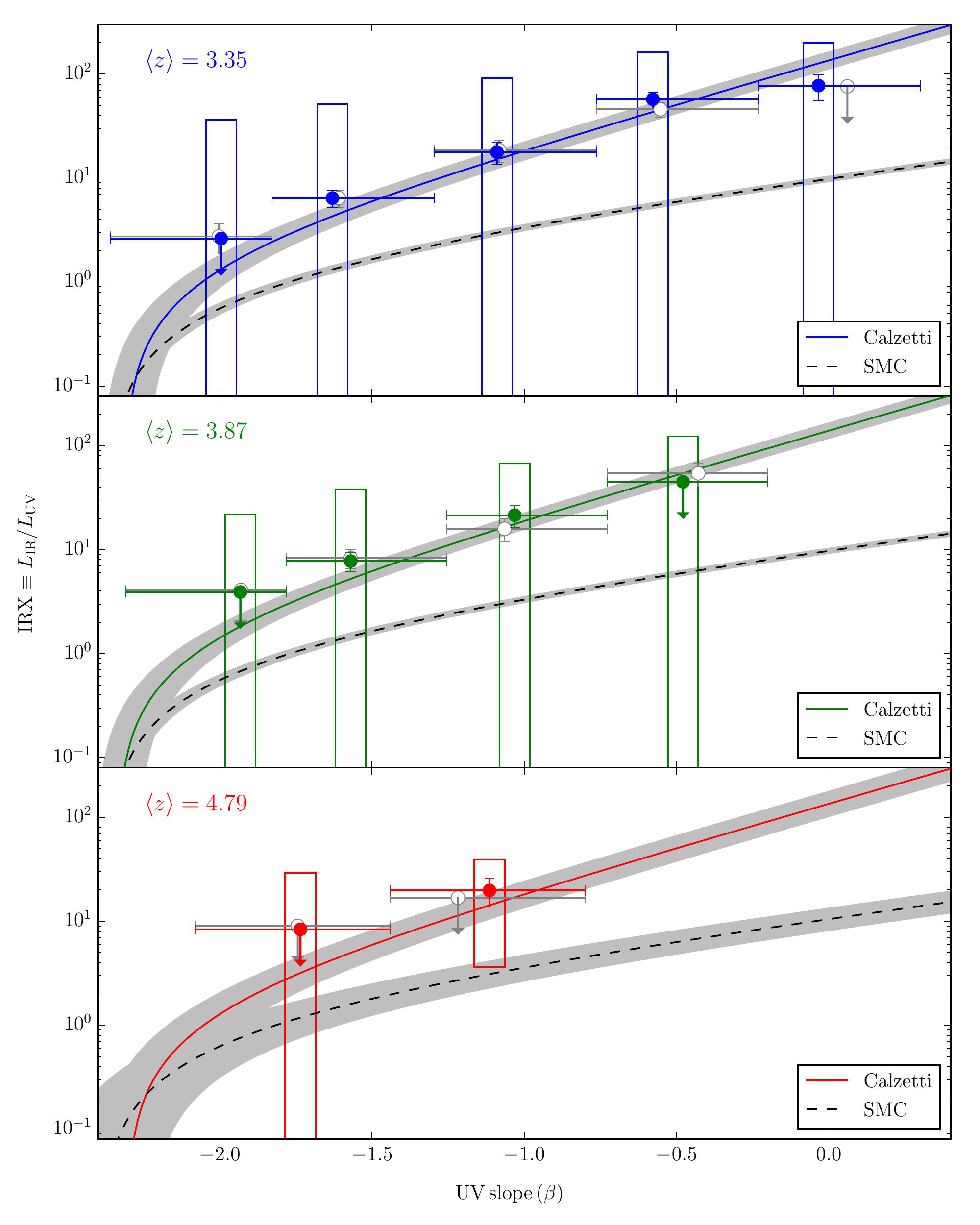}
\end{center}
\caption{IRX-$\beta$ relation for each redshift bin studied in this work. UV slopes were determined from the best-fit SEDs to the rest-frame UV-NIR data only, where the \citet{Calzetti_2000} attenuation curve was used (coloured points). Adopting the SMC-like curve affects the resulting intrinsic (before dust extinction) shape of the stellar emission SED only and has negligible effect on the inferred observed UV slopes. We show this by including data with UV slopes calculated using the SMC-like extinction curve (white points). The coloured points with error bars are the stacked values (Table\,\ref{tab:values}), where we average the IRX values in each $\beta$ and redshift bin (see Section\,\ref{sec:irxs} for details). The bars on $\beta$ merely represent the widths of a given bin, while the values and errors on IRX were found using Equations~\ref{eq:stack} and \ref{eq:stackerr}, respectively (with $3\sigma$ upper limits). The coloured rectangles depict the $1\sigma$ scatter in the individual values of the IRX in each $\beta$ bin. The curves depict the functional forms of the IRX-$\beta$ relation (Table\,\ref{tab:irx}), derived at each redshift bin for Calzetti- and SMC-like dust (see Section\,\ref{sec:irxf} for details). It is clear from this plot that our data are consistent with the Calzetti-like attenuation curve and that there is no obvious redshift evolution of the relation.}
\label{fig:irx}
\end{figure*}

\subsubsection{Stacking IRX}
\label{sec:irxs}

To measure the average ${\rm IRX}\equiv L_{\rm IR}/L_{\rm UV}$ we
first bin the sample in $\beta$. We do not a priori know how $L_{\rm
  IR}$ couples with $L_{\rm UV}$, so we cannot assume that the
$\langle {\rm IRX}\rangle$ in each $\beta$ bin is simply equal to
$\langle L_{\rm IR}\rangle/\langle L_{\rm UV}\rangle$. Therefore we
cannot stack the 850-${\rm \mu m}$ flux densities (i.e. $L_{\rm IR}$)
and divide by the mean $L_{\rm UV}$. Instead, we follow
\citet{Bourne_2017} by assuming that $\langle {\rm IRX}\rangle \equiv
\langle L_{\rm IR}/L_{\rm UV}\rangle$, stacking individual values of
IRX, which is more directly comparable to individually detected
galaxies in the IR (eg. \citealt{Capak_2015, Koprowski_2016}). We find
the individual values of $L_{\rm IR}$ by assuming all LBGs are
described by the average best-fitting template, and normalise this to
the observed 850$\mu$m flux density at the position of each
galaxy. Uncertainties on individual $L_{\rm IR}$ are estimated from
the same measurement in noise-only maps at 850\,${\rm \mu m}$ using
the same scaling factor. The results are presented in
Figure~\ref{fig:irx} and Table\,\ref{tab:values}.

We stress that here the individual values of IRX and $\beta$ have been
calculated independently and that we did not use the energy balance
available in {\sc cigale}. The $L_{\rm IR}$ for each LBG was found
from the best-fit empirical dust-emission SEDs (red curves in
Figure\,\ref{fig:seds}), while the $L_{\rm UV}$ and $\beta$ were
determined from the best-fit SED to the rest-frame UV-NIR photometry
available for each of the sources in our sample. Although {\sc cigale}
uses the \citet{Calzetti_2000} attenuation law, we emphasise that
using the SMC-like extinction curve changes the resulting {\it
  intrinsic} stellar SEDs only, and has negligible effect on the
inferred values of the UV slopes. To illustrate this, we also measure the
observed UV slopes using an SMC-like extinction curve in our {\sc
  cigale} fit (Figure\,\ref{fig:irx}).

\subsubsection{Functional form of IRX-$\beta$ relation}
\label{sec:irxf}

\begin{table}
\begin{footnotesize}
\begin{center}
\caption{Functional forms of the IRX-$\beta$ relation for
  Calzetti-like attenuation and SMC-like extinction curves (see
  Section\,\ref{sec:irxf} for details) plotted in
  Figure\,\ref{fig:irx}.}
\label{tab:irx}
\setlength{\tabcolsep}{2 mm} 
\begin{tabular}{ccc}
\hline
\hline
z    & {\rm IRX}                                  & $A_{1600}$                   \\
\hline
\multicolumn{3}{|c|}{Calzetti-like attenuation curve} \\
\hline
3.35 & $(1.56\pm 0.06)\times(10^{0.4A_{1600}}-1)$ & $2.10(\beta+(2.31\pm 0.07))$ \\
3.87 & $(1.53\pm 0.05)\times(10^{0.4A_{1600}}-1)$ & $2.10(\beta+(2.34\pm 0.07))$ \\
4.79 & $(1.57\pm 0.11)\times(10^{0.4A_{1600}}-1)$ & $2.10(\beta+(2.31\pm 0.10))$ \\
\hline
\multicolumn{3}{|c|}{SMC-like extinction curve} \\
\hline
3.35 & $(1.53\pm 0.04)\times(10^{0.4A_{1600}}-1)$ & $0.92(\beta+(2.37\pm 0.03))$ \\
3.87 & $(1.51\pm 0.03)\times(10^{0.4A_{1600}}-1)$ & $0.92(\beta+(2.37\pm 0.03))$ \\
4.79 & $(1.59\pm 0.13)\times(10^{0.4A_{1600}}-1)$ & $0.92(\beta+(2.39\pm 0.16))$ \\
\hline
\end{tabular}
\end{center}
\end{footnotesize}
\end{table}

We adopt a functional form of IRX from \citet{Meurer_1999}

\begin{equation}\label{eq:irx} {\rm IRX}=(10^{0.4A_{1600}}-1)\times B, \end{equation}

\noindent where $A_{1600}$ is the attenuation at the rest-frame 1600\AA\, in magnitudes and $B$ is the ratio of two bolometric corrections

\begin{equation}\label{eq:bol} B=\frac{\rm BC(1600)}{\rm BC(FIR)}. \end{equation}

\noindent The original \citet{Meurer_1999} relation was defined as ${\rm IRX}\equiv L_{\rm FIR}/L_{\rm UV}$, where

\begin{equation}\label{eq:lfir} L_{\rm FIR}=1.25(L_{60}+L_{100}), \end{equation}

\noindent with $L_{60}$ and $L_{100}$ the luminosities measured by
          {\it IRAS} at 60 and 100\,${\rm \mu m}$. To correct from
          $L_{\rm FIR}$ to total bolometric IR luminosity, the BC(FIR)
          correction was needed. Here we defined IRX as $L_{\rm
            IR}/L_{\rm UV}$, so the IR bolometric correction factor,
          BC(FIR), is by definition equal to unity. The UV bolometric
          correction, BC(1600), converts between all the stellar light
          available to heat the dust and the intrinsic $F_{1600}$
          measured at the rest-frame 1600\AA. This can be calculated
          once the intrinsic stellar emission SED is known by
          integrating between the 912\AA\,(Lyman limit) and
          infinity. As explained above however, to estimate the shape
          of the intrinsic stellar SED one needs to choose the form of
          the attenuation/extinction curve. The difference between the
          two is that the extinction law describes the effects of dust
          in the case of a screen between stars and observer and the
          attenuation law is characteristic of objects in which the
          dust is mixed with stars. Thus, attenuation laws also
          include the effects of scattering of the stellar light into
          our line-of-sight. We consider here two most extreme cases
          of \citet{Calzetti_2000} attenuation curve and SMC-like
          extinction curve \citep{Gordon_2003}.

To find the average intrinsic stellar emission SED corresponding to
each of the attenuation/extinction curves, we used the energy balance
feature of {\sc cigale}, where the amount of the stellar light
attenuated by dust is assumed to be equal to the light re-emitted in
the IR (Table\,\ref{tab:props}). The resulting UV bolometric
corrections, BC(1600), and the intrinsic UV slopes, $\beta_{\rm int}$,
for both attenuation/extinction curves for each redshift bin are given
in Table\,\ref{tab:irx}.

\begin{figure}
\begin{center}
\includegraphics[scale=0.7]{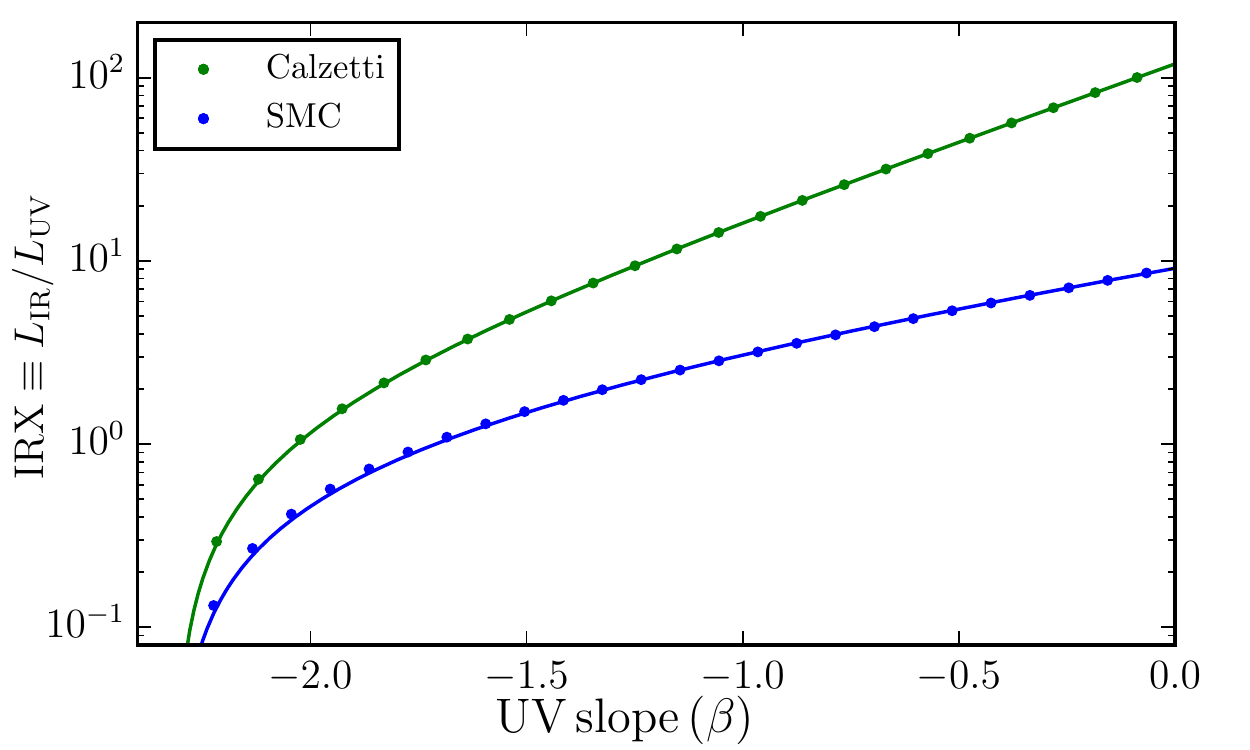}
\end{center}
\caption{Infrared excess as a function of the observed UV slope for Calzetti-like attenuation and SMC-like extinction curves. The points represent the results of the simulations, wherein the unattenuated stellar emission SED was reddened using both, the Calzetti-like attenuation and SMC-like extinction curves in small steps. In each step the resulting IRX value was determined based on the balance between the energy attenuated in the rest-frame UV and re-emitted in the IR. The solid lines are the best-fit functional forms (Equation\,\ref{eq:irx}), with the resulting slopes of 2.1 and 0.92 for the Calzetti- and SMC-like dust, respectively (see Section\,\ref{sec:irxf} for details).}
\label{fig:simm}
\end{figure}

The attenuation at 1600\AA, $A_{1600}$ from Equation\,\ref{eq:irx},
can be described as

\begin{equation}\label{eq:att} A_{1600}=\frac{\delta A_{1600}}{\delta\beta}(\beta_{\rm obs}-\beta_{\rm int}),\end{equation}

\noindent where $\delta A_{1600}/\delta\beta$ is the slope of the
reddening law and $\beta_{\rm obs}$ and $\beta_{\rm int}$ are the
observed and the intrinsic UV slopes, respectively. To find the slope
of the reddening law for the Calzetti- and SMC-like curves, we redden
an intrinsic (dust unattenuated) stellar SED (blue curves in
Figure\,\ref{fig:seds}) in small steps and calculate the amount of the
attenuated stellar light. This is then equated with the energy
re-emitted in the IR by dust. The results of this exercise are
depicted in Figure\,\ref{fig:simm}, where we find slopes of 2.1 for
the Calzetti- and 0.9 for the SMC-like dust.

The resulting functional forms of the IRX-$\beta$ relations
(Equations\,\ref{eq:irx}, \ref{eq:bol} and \ref{eq:att}) for each
attenuation/extinction curve in each redshift bin are summarised in
Table\,\ref{tab:irx} and plotted in Figure\,\ref{fig:irx}. As can be
seen in Figure\,\ref{fig:irx}, assuming different
attenuation/extinction laws does not affect the inferred values of
IRX-$\beta$, which is of course to be expected, since both UV slope
and IRX are observables that should not depend on the model
assumptions. It is therefore clear from Figure\,\ref{fig:irx}, that
our data is consistent with the Calzetti-like dust attenuation,
consistent with \citet{McLure_2017}, and that there is no significant
evolution of the IRX-$\beta$ relation with redshift, as found for the
submm-bright SCUBA-2 galaxies by \citet{Bourne_2017}. It is also clear
from Figure\,\ref{fig:simm}, consistent with the models of
\citet{Narayanan_2017} and \citet{Popping_2017}, that galaxies
following a given IRX-$\beta$ relation have similar stellar
populations and similar intrinsic UV slopes.

\begin{figure*}
\begin{center}
\includegraphics[scale=0.8]{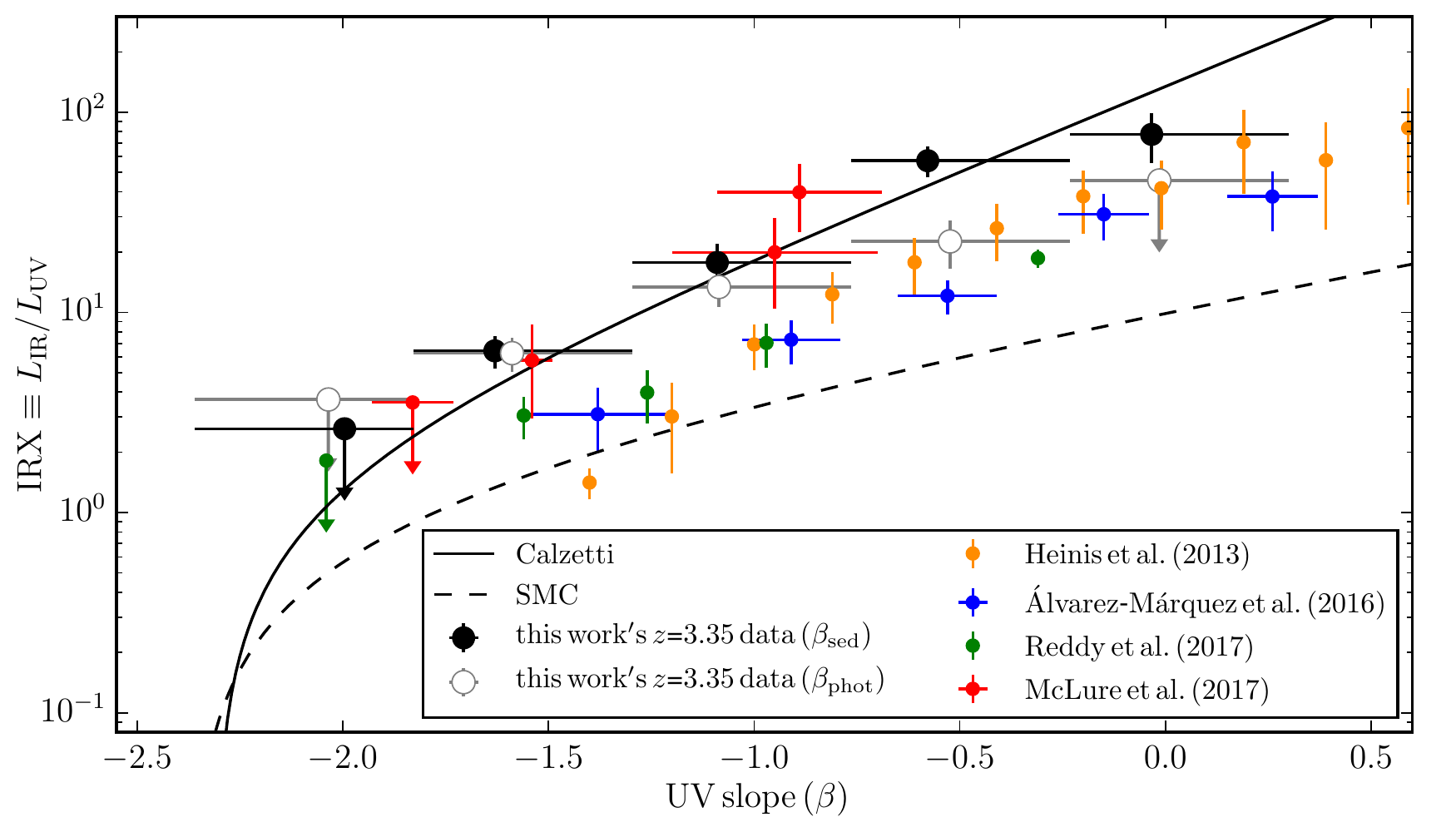}
\end{center}
\caption{IRX-$\beta$ relation for $z\sim 3$ LBGs for this work's sample (black circles), compared with some of the recent literature results \citep{Heinis_2013, Alvarez_2016, Reddy_2017, McLure_2017}. The black solid and dashed lines represent the Calzetti- and SMC-like dust curves from Table\,\ref{tab:irx}. It is clear that, while ours and \citet{McLure_2017} data is consistent with the Calzetti-like dust, others seem to be lying between two dust curves. As shown by \citet{McLure_2017}, this is caused by the uncertainties in the inferred values of the photometry-based UV slopes. We confirm this by including the data with $\beta$'s determined from the best-fit power laws to the rest-frame 1250-2500\AA\, photometry (white circles). One can clearly see that significantly larger errors on the photometry-based values of $\beta$ flatten the slope of the corresponding IRX-$\beta$ relation (see Section\,\ref{sec:comp} for details).}
\label{fig:irxb_comp}
\end{figure*}

\begin{figure*}
\begin{center}
\includegraphics[scale=0.8]{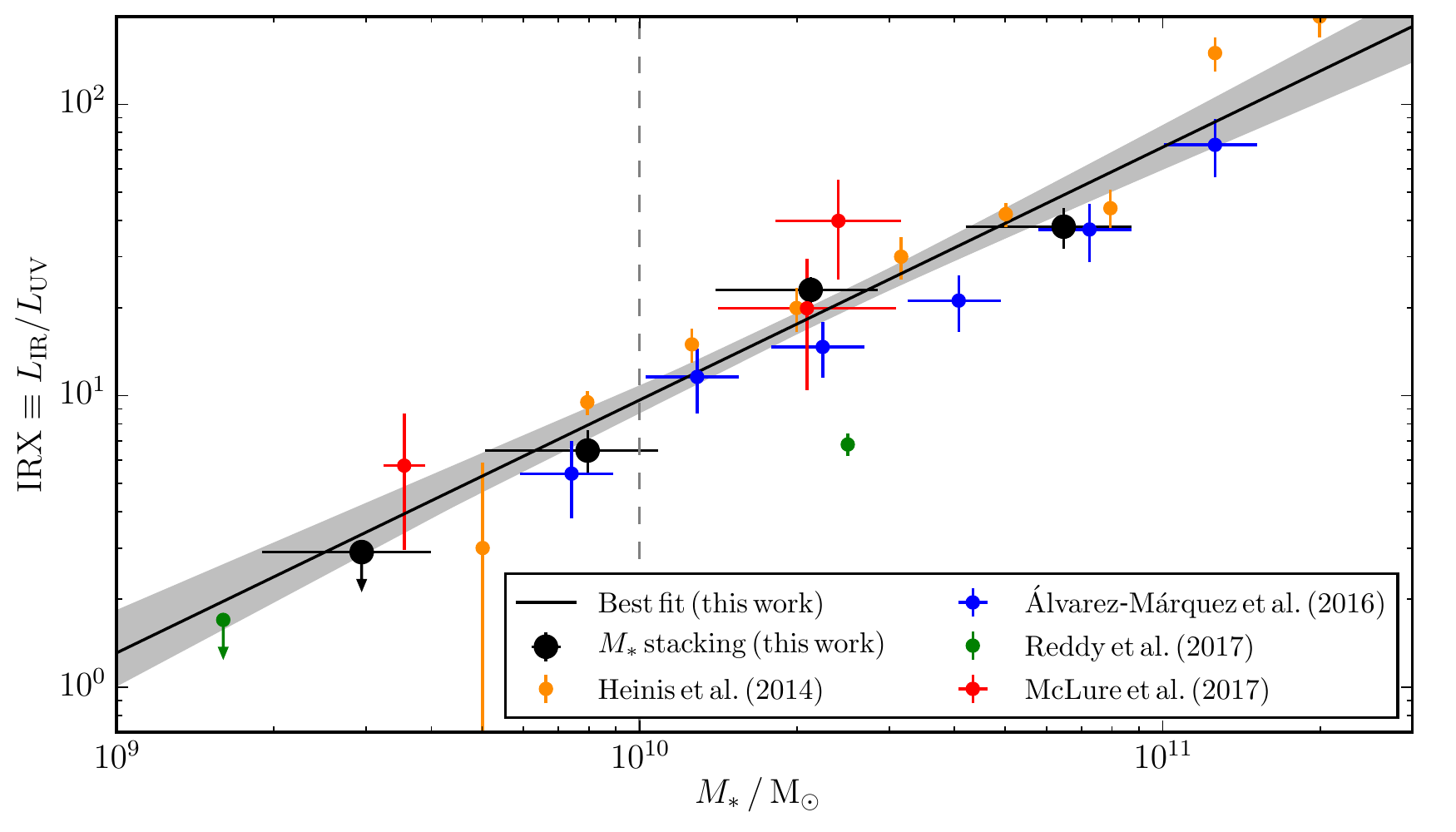}
\end{center}
\caption{IRX-$M_\ast$ relation for our $z\sim 3$ LBGs compared with recent literature results \citep{Heinis_2013, Alvarez_2016, Reddy_2017, McLure_2017} from Figure\,\ref{fig:irxb_comp}. It is clear that the rather striking systematic inconsistencies from Figure\,\ref{fig:irxb_comp} now appear significantly decreased. This further confirms that the scatter in Figure\,\ref{fig:irxb_comp} is mainly driven by different techniques of determining $\beta$. This is because in the IRX-$M_\ast$ relation, stellar masses are determined from the best-fit SEDs, with the corresponding errors of a very similar order (see Section\,\ref{sec:comp} for details). The dashed line represents the mass limit down to which our LBG sample is complete. The lowest-mass upper limit is therefore the only mass-incomplete data point.}
\label{fig:irxm_comp}
\end{figure*}

%We note that there have been a number of individual, direct
%submillimetre/ detections for LBGs at high-$z$ \citep{Capak_2015,
%  Bouwens_2016b, Koprowski_2016}, which do not necessarily follow our
%relation, suggestive of the SMC-like dust law. The thorough
%investigation of those individual IR detections, with the addition of
%the sample of brightest LBGs in the UDS field detected in the ALMA
%Band 7 follow-up observations of SCUBA-2 sources (PI: Smail), will be
%presented in Koprowski et al. (in preparation).

\subsection{Comparison with recent studies}
\label{sec:comp}

In Figure\,\ref{fig:irxb_comp} we compare our $z=3.35$ results with
others works: \citet{Heinis_2013, Alvarez_2016, Reddy_2017} and
\citet{McLure_2017}. Solid and dashed black lines represent the
functional forms of the IRX-$\beta$ relation for Calzetti-like
attenuation and SMC-like extinction curves from
Table\,\ref{tab:irx}. Systematic differences can be immediately noted.
\citet{McLure_2017} and the present work are consistent with
Calzetti-like dust, while other work are intermediate between the
Calzetti- and SMC-like curves. A potential reason for this
inconsistency, pointed out by \citet{McLure_2017} and noted earlier,
is the relatively large uncertainty associated with the determination
of the photometry-based values for UV slopes. Since the reddest
$\beta$ bins are populated by very few sources, a small number of
overestimated UV slopes can cause an apparent drop in IRX, pushing
values towards the SMC-like curve.

To investigate the effects of the scatter of the photometry-based UV
slopes about their real values on the resulting shape of the
IRX-$\beta$ relation, we have re-stacked our $z=3.35$ data. To
estimate the UV slope for each galaxy, we fit a simple power-law to
the photometry available in the rest-frame range of 1250-2500\AA\, and
then stack the IRX in the same $\beta$ bins as in
Section\,\ref{sec:irxs}. The results are shown in
Figure\,\ref{fig:irxb_comp} as white circles. It can be seen that at
the red end, IRX values are suppressed, effectively flattening to
relation and pushing towards the SMC-like curve. This is because, with
our present data, we only have three continuum bands in the rest-frame range
of 1250-2500\AA, resulting in larger errors on $\beta$ and therefore
more scatter in individual $\beta$ bins. Using power-law fits to the
corresponding rest-frame UV range in the best-fitting SEDs, using all
11\,bands of observational data (even if this is not in the nominal
range for a direct estimate of $\beta$) reduces this scatter.

Another approach, taken by \citet{McLure_2017}, is to bin the sample
in stellar mass. This is motivated by the growing consensus that it is
the total stellar mass that influences the amount of the dust
extinction \citep{Heinis_2013, Alvarez_2016, Dunlop_2017,
  Reddy_2017}. We show the stellar mass-binned results of
\citet{McLure_2017} in Figure\,\ref{fig:irxb_comp} as red circles. It
clearly shows, consistent with this work, that $z\sim3$ LBGs are
affected by dust extinction characteristic of the
\citet{Calzetti_2000} law. With $M_\ast$ being a more fundamental parameter,
often the dependence of IRX on $M_\ast$ is determined, instead of UV
slope.  To this end, we stack the $z\approx3$ sample in bins of
$M_\ast$. The results are shown in Figure\,\ref{fig:irxm_comp} as
black circles, with a best-fitting power-law curve of

\begin{equation}\label{eq:irxm} {\rm log(IRX)}=(0.87\pm 0.10)\times {\rm log}(M_\ast/10^{10}{\rm M_\odot}) + (0.98\pm 0.04),\end{equation}

\noindent and the grey area depicting $1\sigma$ uncertainties. Our
results are in excellent agreement with \citet{McLure_2017}, who find
a virtually identical form, with a slope of $0.85\pm 0.05$ and zero
point of $-0.99\pm 0.03$. We also compare to other results in the
literature, corresponding to the data from
Figure\,\ref{fig:irxb_comp}. One can see that the inconsistencies
between different works are much smaller, most likely because the
stellar masses are in all cases determined from the best-fit SEDs
well-sampled with photometry.

\section{Conclusions}
\label{sec:sum}

We have extended the work of \citet{Coppin_2015} to improve on and
calibrate the IRX-$\beta$ relation at $z\simeq 3$-$5$ using 4178
Lyman-break galaxies, stellar mass-complete down to a limit of ${\rm
  log}(M_\ast/{\rm M_\odot})\gtrsim 10.0$. We are able to determine
the average total IR luminosity by stacking galaxies in deep SCUBA-2
850$\mu$m and SPIRE 250--500$\mu$m imaging. By evaulating the observed
UV slope, $\beta$, and emergent UV luminosity, we investigate the
infrared excess, IRX, as a function of observed UV slope and stellar
mass, deriving functional forms. We conclude:

\begin{enumerate}

\item{$3<z<5$ LBGs are consistent with the \citet{Calzetti_2000}
  attenuation law, consistent with the findings of
  \citet{McLure_2017} at $z\sim 3$, now extended to $z\sim 5$. This describes a scenario where dust and stars
  are `well mixed,' on average. In addition, similarly to
  \citet{Bourne_2017}, we find no significant redshift evolution in
  the IRX-$\beta$ over $z\approx3$--$5$.}

\item{the IRX-$\beta$ relationship for LBGs in our sample is
  characteristic of galaxies with similar stellar population ages,
  corresponding to similar intrinsic UV slopes ($\beta_{\rm intr}\sim
  -2.3$), such that observed value of $\beta$ are entirely driven by
  dust obscuration. In turn, this inreases the corresponding IR
  luminosity and hence the IRX. This picture is consistent with the
  theoretical work of \citet{Narayanan_2017} and
  \citet{Popping_2017}.}

 \item{comparing our results with the recent literature findings of
   \citet{Heinis_2013, Alvarez_2016, Reddy_2017} and
   \citet{McLure_2017} we find some inconsistencies, where
   some papers have found significantly lower IRX values for a given $\beta$,
   implying a more `SMC-like' relation. We have confirmed, that these
   inconsistencies are driven by scatter in measured values of $\beta$
   from limited photometry which serves to articifically flatten
   IRX--$\beta$. The scatter is significantly reduced by determining
   $\beta$ from full SED fits, we resolve by determining the values of
   $\beta$ from the best-fit SEDs to the 11 rest-frame UV-NIR
   photometry data points.}

 \item{by stacking IRX in bins of stellar mass, instead of as a
   function of $\beta$ results in a much more consistent
   picture. There is a tight IRX--$M_\ast$ relation in which
   dust-reprocessed stellar emission scales nearly linearly with
   stellar mass. There is much better consistency across different
   works in this parameter space, likely due to the full SED fitting
   used to derive stellar masses, reducing relative uncertainties. We agree that the
   IRX-$M_\star$ relationship is probably a far better proxy for
   correcting observed UV luminosities to total star formation rates,
   provided an accurate handle on $M_\star$ can be had, and also gives
   clues as to the physical driver of dust-obscured star formation in
   high-redshift galaxies.}

\end{enumerate}

\section*{Acknowledgements}

K.E.K.C. and M.P.K. acknowledge support from the UK's Science and Technology Facilities Council (grant number ST/M001008/1). J.E.G. is supported by the Royal Society. M.J.M.~acknowledges the support of the National Science Centre, Poland through the POLONEZ grant 2015/19/P/ST9/04010; this project has received funding from the European Union's Horizon 2020 research and innovation programme under the Marie Sk{\l}odowska-Curie grant agreement No. 665778. KK acknowledge support from the Swedish Research Council and the Knut and Alice Wallenberg Foundation. NB acknowledges support from the European Research Council Advanced Investigator Program, COSMICISM (ERC-2012-ADG-20120216, PI R.J.Ivison).

%%%%%%%%%%%%%%%%%%%%%%%%%%%%%%%%%%%%%%%%%%%%%%%%%%


%%%%%%%%%%%%%%%%%%%%%%%%%%%%%%%%%%%%%%%%%%%%%%%%%%

% Don't change these lines
\bsp	% typesetting comment
\label{lastpage}
\end{document}